\documentstyle[12pt]{article}

\begin{document}

\hspace*{\fill} CERN--TH/95--60\\
\hspace*{\fill} BRX TH--373\\
\hspace*{\fill} ULB--TH--95/04\\

\vspace{3.5cm}
\begin{centering}

{\huge Gauge Properties of Conserved
 Currents in Abelian Versus
 Nonabelian Theories}

\vspace{2cm}
{\large S. Deser$^{1,2}$ and Marc Henneaux$^{3,4}$}\\
\vspace{1cm}
$^1$Physics Department, Brandeis University,
Waltham, MA 02254, USA\\
$^2$CERN,
1211 Geneva 23, Switzerland
\\
$^3$Facult\'e des Sciences, Universit\'e Libre de Bruxelles,
\\Campus Plaine C.P. 231, B-1050  Bruxelles, Belgium\\
$^4$Centro de Estudios Cient\'\i ficos de
Santiago, Casilla 16443, Santiago 9, Chile

\end{centering}

\vspace{2.5cm}

\newpage
\begin{abstract}
We clarify the physical origin of the difference between gauge properties of
conserved currents in abelian and nonabelian theories.  In the latter,
but not in the former, such currents can always be written on shell
as gauge invariants modulo identically conserved, superpotential, terms.
For the ``isotopic" vector and the stress tensor
currents of spins 1 and 2 respectively, we
explain this difference by the fact that the non-abelian
theories are just the self-coupled
versions of the abelian ones using these currents as sources.
More precisely, we indicate how the self-coupling turns
the non-invariantizable abelian conserved currents into (on-shell)
superpotentials. The fate of other conserved currents is also discussed.
\end{abstract}

\newpage\

\section{Introduction}

There is a striking difference between the gauge behaviour of conserved
currents in abelian and nonabelian theories. On the one
hand, it has been recently established \cite{BBH1} for Yang-Mills
theory that (on shell) all local conserved currents either are, or can
be ``improved'' to become, gauge invariant through addition of an identically
conserved
superpotential;  by contrast,  gauge-variant conserved currents \underline{are}
present in the abelian context. On the other hand, it is
known
that one may reach the full nonabelian YM theory as well as Einstein gravity
\cite{SD1}
and
supergravity \cite{DGB} precisely through making the respective original free
abelian
models self-interacting, with their abelian gauge-variant conserved currents as
the sources.
Our aim will be to follow the
abelian to nonabelian
transition to show that these two sets of results are in fact perfectly
complementary.

Physically, one may distinguish two
types of conserved currents: those that are themselves sources --
including possibly of those
fields out of which they are constructed -- and others,
such as the stress tensor
in flat
space physics, that
are not the right-hand sides of any field equations,\footnote{To this category
also
belong the infinitely many conserved ``zilches" -- like the Bel--Robinson
tensor discussed below -- peculiar to free field
theories, but these seem not to be extensible
to the nonabelian regime \cite{DesNic}.}
though their integrated charges may
be physically important.
There is a third category, identically conserved
divergences of superpotentials, and  we shall consider as equivalent two
conserved
currents that differ on-shell by an identically conserved one.\ \ We shall
also say that a current is improvable if it is equivalent on-shell to an
invariant one.

We will show here how the originally non-improvable currents of
the free theories become improved when they  become the (self-coupling) sources
in the nonabelian regime.  We will also
show that all other currents lose their invariance and conservation in the
transition.

\section{Vector Fields}

Consider first the abelian theory, a collection of free photons labelled by a
colour
index; for simplicity of notation, we use the $SU(2)$ case of three such
vectors labelled by (colour)-vector notation, $\mbox{\boldmath $A$}_\mu $. The
gauge-variant Noether current that follows from the invariance of the theory
under
rotations in internal space reads
\begin{eqnarray}
  \mbox{\boldmath $j$}^\mu    & = & \mbox{\boldmath $F$}^{\mu \nu }_{abel}
\times \mbox{\boldmath $A$}_\nu , \;
  \mbox{\boldmath $F$}^{\mu \nu }_{abel}    =
\partial _\mu \mbox{\boldmath $A$}_\nu -\partial _\nu \mbox{\boldmath $A$}_\mu
   \label{current}
\end{eqnarray}
and is manifestly conserved on shell ($\partial _\mu
\mbox{\boldmath $F$}_{abel}^{\mu \nu }=0$). It is not improvable to be
invariant under the abelian gauge transformations of the free theory (even on
shell),
being bilinear in the fields but involving only a single derivative. As
indicated in \cite{BBH1},
it is also the only non-improvable conserved current on a generic background
spacetime.

Now, if we make $\mbox{\boldmath $j$}^\mu $ the source of its own field
\begin{eqnarray}
  \partial _\nu \mbox{\boldmath $F$}^{\mu \nu }   & = &
  \mbox{\boldmath $j$}^\mu ,
  \label{YMequ}
\end{eqnarray}
(with the usual rescaling of the self-coupling constant)
then antisymmetry of $\mbox{\boldmath $F$}^{\mu \nu }$ requires
$\mbox{\boldmath $j$}^\nu $ to be conserved also on the new mass shell.  A
simple
calculation  \cite{SD1} shows that this in turn requires the $\mbox{\boldmath
$F$}%
^{\mu \nu }$ in (\ref{YMequ}) to be promoted to its full Yang-Mills form,
\begin{equation}
\mbox{\boldmath $F$}^{\mu \nu } =  \mbox{\boldmath $F$}^{\mu \nu }_{ym} =
\mbox{\boldmath $F$}_{abel}^{\mu \nu }+
  \mbox{\boldmath $A$}_\mu \times \mbox{\boldmath $A$}_\nu ,
\label{Fieldstrength}
\end{equation}
so that (\ref{YMequ}) become the usual Yang-Mills equations
$\mbox{\boldmath $D$}_\mu \mbox{\boldmath $F$}_{ym}^{\mu \nu }
= (\partial_\mu + \mbox{\boldmath $A$}_\mu \times )$ \linebreak
$\mbox{\boldmath
$F$}^{\mu\nu}_{ym}
=0$. The abelian current in (\ref
{current}) is correspondingly promoted to be a conserved gauge invariant object
under the
non-abelian deformation of the original abelian transformation.\ \ This
promotion is formally trivial; just write\footnote{One could also define a
seemingly
different $\mbox{\boldmath $J$}^\mu$ as the right side of
$\partial_\mu \mbox{\boldmath $F$}^{\mu\nu}_{abel} = \mbox{\boldmath $J$}^\nu$,
but it merely differs from
$\mbox{\boldmath $j$}^\mu $ by the superpotential
$\partial_\mu (\mbox{\boldmath $A$}_\mu \times \mbox{\boldmath $A$}_\nu)$.}
\begin{equation}
\mbox{\boldmath $j$}^\mu =
\mbox{\boldmath $F$}^{\mu\nu}_{ym} \times \mbox{\boldmath $A$}_\nu =
-\mbox{\boldmath $D$}_\nu \mbox{\boldmath $F$}_{ym}^{\mu \nu }+
\partial_\nu \mbox{\boldmath $F$}_{ym}^{\mu \nu },
\label{current2}
\end{equation}
to exhibit that the current is the sum of a superpotential and a gauge
covariant term; the latter term actually vanishes on-shell, in agreement with
the general theorem of \cite{BBH1} according to which a conserved current in
Yang-Mills theory can always be redefined to be gauge {\em invariant} by the
addition of equation of motion and superpotential terms. In the abelian
limit, the right-hand side of (\ref{current2}) would of course identically
vanish, reminding us of the non-improvability there. [Note also that $%
\mbox{\boldmath $j$}^\mu $ is {\em not} covariantly conserved, due to the
superpotential in (\ref{current2}).]
Thus we see that the consistent self-coupling of massless vector fields
provides a mechanism by which the non-improvable current (\ref{current}) of
the abelian theory becomes improved and actually equivalent to an
identically conserved current. This follows from the precise form of the
interaction, which promotes the originally non-improvable current to be the
source of the divergence of the antisymmetric tensor $\mbox{\boldmath $F$}%
_{ym}^{\mu \nu }$. Differently put, the current $\mbox{\boldmath $j$}^\mu $ is
improvable as soon as there is a shell for it to be on. This shell is given
by the consistent self-coupling, which eliminates all the non-improvable
currents.

The abelian theory also contained one (and only one) conserved (and invariant)
tensor
current, namely the field strength $\mbox{\boldmath $F$}_{abel}^{\mu \nu}$
itself \cite{BBH2}.
Corresponding to the fact that $\mbox{\boldmath $F$}^{\mu \nu}$ becomes
covariant in
the nonabelian case, it also becomes covariantly conserved, so that there
remain no ordinarily conserved
higher rank quantities either.

For completeness, we consider finally the rather different, identically
conserved
currents such as the abelian
\begin{eqnarray}
\tilde{\mbox{\boldmath ${\jmath}$}}^\mu_{abel} =  \mbox{\boldmath $A$}_\nu
\times \tilde{\mbox{\boldmath ${F}$}}_{abel}^{\mu \nu }, \;\;
\equiv \textstyle{\frac {1}{2}}\; \partial_\alpha \;
\epsilon^{\mu\nu\alpha\beta} \mbox{\boldmath ${A}$}_\nu \times
\mbox{\boldmath ${A}$}_\beta \; ; \;\;\;\;\;
\tilde{\mbox{\boldmath ${F}$}}_{abel}^{\mu \nu }
\equiv \textstyle{\frac {1}{2}}\; \epsilon ^{\mu \nu \alpha \beta }
\mbox{\boldmath $F$}_{\alpha \beta }^{abel}
\label{dual}
\end{eqnarray}
Although the non-abelian generalization does {\em not} promote
$\tilde{\mbox{\boldmath $\jmath$}}^\mu_{abel}$ to be a source current,
$\tilde{\mbox{\boldmath $\jmath$}}^\mu_{abel}$ remains identically conserved
nevertheless if (as was
done for $\mbox{\boldmath $j$}^\mu$ above)
$\tilde{\mbox{\boldmath $F$}}^{\mu \nu }_{abel}$
is replaced by
$\tilde{\mbox{\boldmath $F$}}^{\mu\nu}_{ym}$,
as is easily verified using the Bianchi identity ${\mbox{\boldmath $D$}}_\mu
\tilde{\mbox{\boldmath $F$}}_{ym}^{\mu \nu }\equiv 0$. Indeed, here too we can
write the superpotential form
\begin{equation}
   \tilde{\mbox{\boldmath $\jmath$}}^\nu_{ym}    =  0 +
\partial_\mu ^{}\tilde{\mbox{\boldmath $F$}}_{ym}^{\mu \nu }
\label{superbis}
\end{equation}
as an identity, so although
$\tilde{\mbox{\boldmath $\jmath$}}^\mu_{ym} = \tilde{\mbox{\boldmath $\jmath$}}
^\mu_{abel}$, it is
nonabelian gauge invariant (namely zero) up to a superpotential, quite
independently of any Yang-Mills dynamics.

In d=3,  one may also consider the Chern-Simons action
$I = \int d^3 x \mbox{\boldmath $A$}_\mu \cdot \tilde{\mbox{\boldmath
$F$}}^\mu_{abel}$; the corresponding abelian current is
the ``descent" of (\ref{dual}) to $d=3$, namely
$\mbox{\boldmath $j$}^{\mu}_{abel} = \epsilon^{\mu\alpha\beta}  \mbox{\boldmath
$A$}_\alpha \times \mbox{\boldmath $A$}_\beta$, conserved
by virtue of the field equations $\tilde{\mbox{\boldmath $F$}}^\mu_{abel} = 0$.
Unlike (\ref{dual}), this $\mbox{\boldmath $j$}^\mu$ \underline{not}
identically conserved and so \underline{can} be set as a
source of $\tilde{\mbox{\boldmath $F$}}^\mu$.  The resulting  non-abelian field
equation, $\mbox{\boldmath $F$}^{\mu\nu}_{ym} = 0$,  implies that
$\mbox{\boldmath $j$}^\mu$ is an on-shell gauge invariant object,
\begin{equation}
\mbox{\boldmath $j$}^\mu = \tilde{\mbox{\boldmath $F$}}^\mu
+ 2 \partial_\alpha \;
\epsilon^{\mu\alpha\beta}\mbox{\boldmath $A$}_\beta
\end{equation}
just as in (\ref{current2}).

\section{Tensor Fields}

The spin 2 situation is so similar to that for spin 1 that we omit all details
(see [2]).  The
analog of the iso-current (1) is the stress tensor $T^{\mu\nu}$ of the free
field action, which
is conserved but neither gauge invariant nor improvable, being of second
derivative order,
while invariance would require it to be bilinear in the (linearized) curvature
and hence of
fourth order.  When $T^{\mu\nu}$ is made the source of the linearized Einstein
equations, thereby defining the full Einstein theory,
it remains
(ordinarily, not covariantly)
conserved -- on full mass shell -- and can be expressed, in parallel
with (\ref{current2}), as
\begin{eqnarray}
T^{\mu \nu}   & = &  - G^{\mu \nu} + G^{\mu \nu}_L
\label{stress2}
\end{eqnarray}
where $G^{\mu \nu }$ is the full covariant Einstein tensor
and its linearization is of course the divergence of a superpotential,
$G^{\mu\nu}_L = \partial_\lambda S^{\lambda\mu\nu}$. Thus
$T^{\mu\nu}$ is itself  deformed gauge (diffeomorphism) invariant on shell up
to that superpotential.

In the vector case, we noted the existence of the linear tensor current,
$\mbox{\boldmath $F$}^{\mu\nu}$, conserved (only) at abelian level.  There are
actually two different higher rank analogs here.  The first is linear in the
fields, like $\mbox{\boldmath $F$}^{\mu\nu}$: it is the set of superpotentials
$S^{\lambda\mu\nu}$ of $G^{\mu\nu}_L$.  They are conserved on linear Einstein
shell, but (unlike $\mbox{\boldmath $F$}^{\mu\nu}$) gauge-variant; their
conservation (like that of $\mbox{\boldmath $F$}^{\mu\nu}$) does not survive
self-coupling \cite{BBH2}.  The second higher rank quantity is more akin to the
Maxwell stress-tensor, being bilinear in the fields:
The abelian spin 2 theory allows for the existence
of the gauge-invariant, conserved, totally symmetric and traceless (in d=4),
Bel--Robinson
tensor.\footnote{This fourth-rank tensor cannot, however, be made the source of
any gravitational quantity nor even of a spin 4 field \cite{6prime}.}
\begin{equation}
B_{\alpha\beta\gamma\delta} = R_{\alpha\rho\gamma\sigma} \;
R_\beta~\!\!^\rho~\!\!_\delta~\!\!^\sigma +
\tilde{R}_{\alpha\rho\gamma\sigma}\;
\tilde{R}_\beta~\!\!^\rho~\!\!_\delta~\!\!^\sigma \; ,
\end{equation}
where $\tilde{R}$ is the (left pair) dual.  [Compare the vector field's stress
tensor, $2T_{\mu\nu} = \mbox{\boldmath $F$}_\mu ~^\alpha \cdot
\mbox{\boldmath $F$}_{\nu\alpha} + \tilde{\mbox{\boldmath $F$}}_{\mu}~^\alpha
\cdot
\tilde{\mbox{\boldmath $F$}}_{\nu\alpha}$.]  Just as $\mbox{\boldmath
$F$}_{\mu\nu}$ does,
$B_{\alpha\beta\gamma\delta}$ becomes
gauge-{\em covariant} and {\em covariantly} conserved in the full
theory; there are no ordinarily conserved diffeomorphism-invariant local
quantities in curved space.

\section{Comments}

The two examples we have exhibited of gauge-variant current improvement through
the mechanism of self-coupling are essentially unique.  [The only other case is
the construction of supergravity \cite{DGB} using the simultaneous
gravitational stress tensor self coupling and that of the spin 3/2--spin 2
supercurrent.]  Gauge fields of higher spin just do not possess local conserved
free-field currents of sufficiently high rank to replicate the constructions
given above. Spin 3 provides the simplest example of this failure \cite{TD}.
Of course, using a conserved abelian current is not the only consistent way to
self-couple a gauge system, but it is the only physically interesting one
\cite{Wald}.

In itself, the existence of gauge variant conserved currents in abelian
theories is not worrisome, so long as they are not physical sources.  For
example, as we have noted, the
stress tensors of any spin $>$2 gauge fields are non-improvable, but they do
obey the one physical requirement that their integrated charges -- the Poincare
generators -- be gauge invariant \cite{SD2}.
Indeed, those free fields -- (abelian) N-forms  -- that do have invariant
stress tensors, instead
encounter
a clash between conservation and gauge invariance (in all but one critical
dimension) when it comes to the conformal group \cite{SDSch}. There is no
local Lorentz covariant expression for even the dilation current that
respects gauge invariance, as witness the unimprovable
expression \cite{BBH1} for
Maxwell theory,
\begin{eqnarray}
   D^\mu   & = & x^\nu T^\mu_\nu +
\textstyle{\frac{1}{2}} \;(d-4)F^{\mu \nu}_{abel}  A_\nu.
\label{dilation}
\end{eqnarray}
Expressions similar to (\ref{dilation}) hold for arbitrary N-form fields
\cite{SDSch} in dimension other than $2(N+1)$, and indeed conformal (but of
course not Poincar\'e!) charges lose their significance as a result.

In summary, we have provided a physical explanation of the findings of
\cite{BBH1,BBH2}
about the existence of non-improvable gauge variant currents in free spin 1 and
2 gauge
theories and their absence when consistent self-interactions are turned on.
In particular, we saw that
those currents that can be turned into self-sources thereby acquire the --
deformed -- gauge covariance of the resulting nonlinear theory, modulo
superpotentials. [Identically conserved currents behave similarly without use
of
dynamics.] For higher
spins there is no self-coupling mechanism available, and even N-form abelian
fields have ``bad'' conserved currents that cannot generally be improved in a
local fashion.

\section{Acknowledgements}

M.H. is grateful to Glenn Barnich and Friedemann Brandt for
useful conversations. This work has been supported in part by
NSF grant PHY--9315811, by
research
contracts with the Commission of the European Communities and by research
funds from the Fonds National de la Recherche Scientifique (Belgium).

\newpage\

\end{document}